# An Objective Vortex Identification Method

Yifei Yu[1], Yiqian Wang[2] and Chaoqun Liu[1][†]
1. Department of Mathematics, University of Texas at Arlington, Arlington, Tx, USA, 76019
2. School of Mathematical Science, Soochow University, Suzhou 215006, China

Abstract:
Generally, the vortex structures should be independent of the observers who are moving, especially when their coordinates are non-inertial which may result in confusions in communications between researchers. The property that not being influenced by the choice of coordinate is called objective. Mostly, researchers would like to gather data in the inertial coordinate because most physical laws are valid in the inertial coordinate. As a result, how to transfer the observations in a non-inertial coordinate to those in an inertial coordinate raises people's interest. In this paper, methods are provided to obtain velocity and velocity gradient tensor in an inertial coordinate from the non-inertial observations based on one known zero-vorticity point which can be chosen from physical fact such as the point is located in the inviscid region. Further, objective vortex structure can be calculated from the objective velocity or velocity gradient tensor. With Liutex chosen as the indicator of the vortex, two numerical examples are used to test the proposed methods. The results show that the methods perform very well and the relative error is at most at the order of $10^{-7}$.

1. Introduction

Vortex is an important element in fluid dynamics research. As aptly remarked by Kiichemann (Küchemann 1965), vortices are the 'sinews and muscles of fluid motions'. Scholars' understanding of vortex has gone through many stages. This first stage is vorticity and vorticity-related methods. Vortex known as the flow revolves around an axis, was first represented by vorticity because vorticity represents the rotation of a rigid body. However, the vorticity may contain part of the shear when applied to fluid since ideal rigid bodies do not have any deformations while fluid cannot resist shear without any deformations. For flows with small shear, vorticity can detect vortex approximately, but it does not match the experimental results for flows with strong shears. For example, in the near wall region of the turbulence, Robinson(Robinson 1991) reported the relation between vorticity and vortex is very weak. Yu (Yu et al. 2021) analyzed how vorticity is contaminated by shear and provided the correlation between vorticity and vortex from lower boundary layer to the upper boundary layer. To overcome the drawback of using vorticity, scientists started to develop some new methods which are the second-generation methods. The second-generation methods are dependent on velocity gradient tensor such as Q criterion (Hunt, Wray, and Moin 1988), $\lambda_{ci}$ criterion (Zhou et al. 1999), $\lambda_2$ criterion (Jeong and Hussain 1995), $\Delta$ criterion (Chong, Perry, and Cantwell 1990) and etc. Although these methods match the experimental results better than vorticity, they suffer from some common problems. First, all second-generation methods are scalar without the direction of the swirling axis. Second, the physical meaning of the numbers given by these methods is not clear. Admittedly, the values of these methods reflect the relative rotation strength, people do not know the relation between these values and the accurate angular speed. An ideal vortex identification method needs to indicate the accurate angular speed rather than the approximate rotation strength. Liu proposed Liutex (Liu et al. 2018), a third-generation method. Liutex is a vector physical quantity with a clear physical meaning. The direction of Liutex is the swirling axis and the magnitude of Liutex is the twice angular speed. Even though Liutex is a new-born method, it has been used and tested by many researchers. Liang et al. (Liang et al. 2022) used Liutex to analyze the interaction between cavitation and vortex. Guo et al. (Guo et al. 2020) compare Liutex with other vortex identification methods based on a vortex evolution experimental result. Zhao et al. (Zhao, Wang, and Wan 2020) applied Liutex in marine hydrodynamics. Some other researchers also made some contributions to the Liutex method (Shrestha et al. 2021; Yu et al. 2021; Dong and Liu 2021; Wang et al. 2019; Zhao et al. 2021). In this paper, Liutex is used as the vortex identification method to detect vortex.

Objectivity is a property that the observed results keep the same under different coordinates. Objectivity is of great importance in vortex detection because different people who may choose different coordinates, and if the results are different under different coordinates, then people do not have a uniform standard to compare the experimental results done by different researchers standing in different coordinates. Take the streamline as an example. In Fig. 1, the streamline looks like a straight line and no vortex can be observed. But if the coordinate is changed by subtracting the u-velocity at the selected point, i.e. the coordinate is moving at the u-velocity, which is the velocity component at the selected point, then obvious flow rotation can be seen. Unfortunately, most of the vortex identification methods are not objective. To avoid the confusion caused by the coordinate changes, many scholars devote themselves to find an objective vortex identification method. A natural idea is to find a distinguished frame of reference (usually called co-moving or proper frame (Landau 2013)). For the flows that have a homogeneous direction of the flow propagation, the co-moving frame can be found (Waleffe 2001; Mellibovsky and Eckhardt 2012; Kreilos, Zammert, and Eckhardt 2014). Nevertheless, this idea does not seem feasible for general fluid flows, as pointed out by Lugt (Lugt 1979). Another idea is to give a new objective definition to vortex. In 2005, Haller (Haller 2005) defined "*a vortex is a set of fluid trajectories along which $M_Z$ is indefinite*". Although this definition is objective, it does not reflect the rotation strength. Some other researchers aim to modify the existing vortex identification methods to be objective. For example, there are a number of reports (Martins et al. 2016; Günther, Gross, and Theisel 2017; Hadwiger et al. 2018; Liu, Gao, and Liu 2019; Rojo and Gunther 2020).

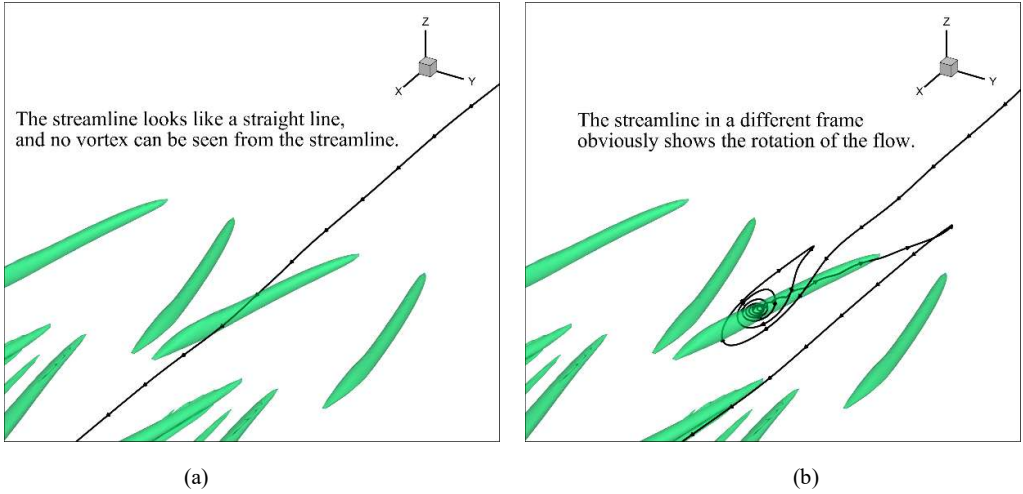

(a) (b)

Fig. 1 Streamline in (a) original coordinate (b) the observer moves at the same u-velocity of the selected point.

In this paper, the objective velocity and velocity gradient tensor are derived for the situation that zero vorticity points can be found based on the physical properties. Using one zero vorticity point as the reference point, the velocity field and the velocity gradient tensor field can be calibrated to the results under an inertial coordinate. In addition, Liutex is Galilean invariant (Liu 2018), which means it keeps identical under different inertial coordinates. Therefore, an objective Liutex called $\vec{L}_{ob}$ can then be obtained. Any vortex identification methods that are Galilean invariant can use this method to become objective by using the objective velocity or the objective velocity gradient tensor.

This paper is organized as follows. Section 2 and 3 introduce the basic knowledge of objectivity and Liutex briefly. The methods to calculate objective vortex from non-inertial observations are provided in Section 4 with detailed derivation. In Section 5, two numerical examples are carried out to test the effectiveness of the methods. Then, conclusion is given in Section 6.

2.  Objectivity and Objective Results

According to the classic theorem, a tensor field $A(\vec{x},t)$ is objective (Gurtin 1982; Gurtin, Fried, and Anand 2010) if its corresponding tensor field $\tilde{A}(\vec{y},t)$ satisfies

$$\tilde{A}(\vec{y},t) = Q^T(t)A(\vec{x},t)Q(t) \quad (2.1)$$

after the Euclidean frame change
$$\vec{x} = Q(t)\vec{y} + b(t) \tag{2.2}$$
where $Q(t)$ is any rotation tensor and $b(t)$ is any translation vector.
A vector $\vec{x}$ is objective if its corresponding vector $\vec{y}$ satisfies
$$\vec{x} = Q^T(t)\vec{y} \tag{2.3}$$
after the Euclidean frame change
$$\vec{x} = Q(t)\vec{y} + b(t) \tag{2.4}$$
where $Q(t)$ is any rotation tensor and $b(t)$ is any translation vector.

The purpose of this paper is to derive a method to get objective vortex structure. Objective vortex means the same quantity and local rotation axis direction of vortex can be drawn by different observers who are standing in different inertial or non-inertial coordinates. For example, let any two observers observe the same flow, one is stationary while the other is moving. If the same vortex structure can be obtained from these two observations, then we say this vortex structure is objective.

3.  Liutex

Liutex $\vec{R} = R\vec{r}$ is a vector whose direction $\vec{r}$ is the real eigenvector of the velocity gradient tensor with $\vec{\omega} \cdot \vec{r} > 0$ where $\vec{\omega}$ is the vorticity, and whose magnitude $R$ can be calculated by
$$R = \vec{\omega} \cdot \vec{r} - \sqrt{(\vec{\omega} \cdot \vec{r})^2 - 4\lambda_{ci}^2} \tag{3.1}$$
where $\lambda_{ci}$ represents the real part of the conjugate complex eigenvalues of the velocity gradient tensor.

The direction of Liutex refers to the local rotation axis and its magnitude is twice angular speed.

4.  Obtain Objective Vortex Results from One Known Zero-Vorticity Point

Most physical laws are valid under inertial coordinates, so the objective results in this section refer to the result under an inertial coordinate. To find the objective Liutex result under an inertial coordinate, people only need to find the Liutex in any inertial coordinate since Liutex is Galilean invariant, which is consistent under different inertial coordinates. The rest of this section derives a method to calculate objective Liutex in an inertial coordinate based on observations in a non-inertial coordinate.

Upper case letters are used to express variables in the inertial coordinate. Let $X, Y, Z$ be the position components; $\vec{X}, \vec{Y}, \vec{Z}$ be the unit basis vectors of the coordinate; $U, V, W$ be velocity components; $\vec{V}, \nabla\vec{V}$ be velocity vector and velocity gradient tensor. Lower case letters are used to express variables in the observer's (non-inertial) coordinate. Let $x, y, z$ be the position components; $\vec{x}, \vec{y}, \vec{z}$ be the unit basis vectors of the observer's coordinate; $u, v, w$ be velocity components; $\vec{v}, \nabla\vec{v}$ be velocity vector and velocity gradient tensor. The velocity of the observer in the inertial coordinate is $\vec{V}_{ob} = \vec{V}_t + \vec{V}_a$, where $\vec{V}_t$ is the translation velocity and $\vec{V}_a = V_{ax}\vec{X} + V_{ay}\vec{Y} + V_{az}\vec{Z}$ is the angular velocity respectively. Note that, in this paper, $\vec{V}_a$ does not represent velocity but angular speed. The observer can be regarded as a rigid point, so $\vec{V}_{ob}$ is only dependent on time $t$. According to the knowledge of kinetics, $\vec{V}$ and $\vec{v}$ have the following relation,
$$\vec{V}(\vec{P}) = \vec{v}(\vec{p}) + \vec{V}_t(t) + \vec{V}_a(t) \times \vec{P} \tag{4.1}$$

where $\vec{P}$ is the position vector of the observer in the inertial coordinate and $\vec{p}$ is the position vector of the observer in the observer's coordinate. Use the unit basis vectors to express $\vec{P}$, $\vec{p}$, $\vec{V}(\vec{P})$ and $\vec{v}(\vec{p})$

$$\vec{P} = X\vec{X} + Y\vec{Y} + Z\vec{Z} \tag{4.2}$$
$$\vec{p} = x\vec{x} + y\vec{y} + z\vec{z} \tag{4.3}$$
$$\vec{V}(\vec{P}) = U(\vec{P})\vec{X} + V(\vec{P})\vec{Y} + W(\vec{P})\vec{Z} \tag{4.4}$$
$$\vec{v}(\vec{p}) = u(\vec{p})\vec{x} + v(\vec{p})\vec{y} + w(\vec{p})\vec{z} \tag{4.5}$$

Suppose the transition matrix from $\vec{x}, \vec{y}, \vec{z}$ to $\vec{X}, \vec{Y}, \vec{Z}$ is $T$, i.e.

$$[\vec{x} \quad \vec{y} \quad \vec{z}] = [\vec{X} \quad \vec{Y} \quad \vec{Z}]T \tag{4.6}$$

where

$$T = \begin{bmatrix} \cos\alpha_1 & \cos\alpha_2 & \cos\alpha_3 \\ \cos\beta_1 & \cos\beta_2 & \cos\beta_3 \\ \cos\gamma_1 & \cos\gamma_2 & \cos\gamma_3 \end{bmatrix} \tag{4.7}$$

$\cos\alpha_1, \cos\beta_1, \cos\gamma_1$ are the direction cosines of $\vec{x}$ in XYZ-coordinate; $\cos\alpha_2, \cos\beta_2, \cos\gamma_2$ are the direction cosines of $\vec{y}$ in XYZ-coordinate; $\cos\alpha_3, \cos\beta_3, \cos\gamma_3$ are the direction cosines of $\vec{z}$ in XYZ-coordinate.

From Eq. (4.6),

$$\vec{x} = \cos\alpha_1 \vec{X} + \cos\beta_1 \vec{Y} + \cos\gamma_1 \vec{Z} \tag{4.8}$$

Thus,

$$\frac{\partial \vec{x}}{\partial \vec{X}} = \cos\alpha_1 \tag{4.9}$$

Similarly,

$$\begin{bmatrix} \frac{\partial \vec{x}}{\partial \vec{X}} & \frac{\partial \vec{y}}{\partial \vec{X}} & \frac{\partial \vec{z}}{\partial \vec{X}} \\ \frac{\partial \vec{x}}{\partial \vec{Y}} & \frac{\partial \vec{y}}{\partial \vec{Y}} & \frac{\partial \vec{z}}{\partial \vec{Y}} \\ \frac{\partial \vec{x}}{\partial \vec{Z}} & \frac{\partial \vec{y}}{\partial \vec{Z}} & \frac{\partial \vec{z}}{\partial \vec{Z}} \end{bmatrix} = \begin{bmatrix} \cos\alpha_1 & \cos\alpha_2 & \cos\alpha_3 \\ \cos\beta_1 & \cos\beta_2 & \cos\beta_3 \\ \cos\gamma_1 & \cos\gamma_2 & \cos\gamma_3 \end{bmatrix} \tag{4.10}$$

Eq. (4.1) can be written as

$$[\vec{X} \quad \vec{Y} \quad \vec{Z}] \begin{bmatrix} U(\vec{P}) \\ V(\vec{P}) \\ W(\vec{P}) \end{bmatrix} = [\vec{x} \quad \vec{y} \quad \vec{z}] \begin{bmatrix} u(\vec{p}) \\ v(\vec{p}) \\ w(\vec{p}) \end{bmatrix} + [\vec{X} \quad \vec{Y} \quad \vec{Z}] \begin{bmatrix} V_{tx} \\ V_{ty} \\ V_{tz} \end{bmatrix} + [\vec{X} \quad \vec{Y} \quad \vec{Z}] \begin{bmatrix} V_{ay}Z - V_{az}y \\ V_{az}X - V_{ax}Z \\ V_{ax}Y - V_{ay}X \end{bmatrix} \tag{4.11}$$

Substitute $[\vec{x} \quad \vec{y} \quad \vec{z}]$ using Eq. (4.6)

$$[\vec{X} \quad \vec{Y} \quad \vec{Z}] \begin{bmatrix} U \\ V \\ W \end{bmatrix} = ([\vec{X} \quad \vec{Y} \quad \vec{Z}]T) \begin{bmatrix} u \\ v \\ w \end{bmatrix} + [\vec{X} \quad \vec{Y} \quad \vec{Z}] \begin{bmatrix} V_{tx} \\ V_{ty} \\ V_{tz} \end{bmatrix} + [\vec{X} \quad \vec{Y} \quad \vec{Z}] \begin{bmatrix} V_{ay}Z - V_{az}y \\ V_{az}X - V_{ax}Z \\ V_{ax}Y - V_{ay}X \end{bmatrix} \tag{4.12}$$

Take partial derivative with respect to $X$ on both sides of Eq. (4.12)

$$\frac{\partial}{\partial X}[\vec{X} \quad \vec{Y} \quad \vec{Z}] \begin{bmatrix} U \\ V \\ W \end{bmatrix} = \frac{\partial}{\partial X} \left\{ [\vec{X} \quad \vec{Y} \quad \vec{Z}]T \begin{bmatrix} u \\ v \\ w \end{bmatrix} + [\vec{X} \quad \vec{Y} \quad \vec{Z}] \begin{bmatrix} V_{tx} \\ V_{ty} \\ V_{tz} \end{bmatrix} + [\vec{X} \quad \vec{Y} \quad \vec{Z}] \begin{bmatrix} V_{ay}Z - V_{az}y \\ V_{az}X - V_{ax}Z \\ V_{ax}Y - V_{ay}X \end{bmatrix} \right\} \tag{4.13}$$

$\frac{\partial V_{tx}}{\partial X} = \frac{\partial V_{ty}}{\partial X} = \frac{\partial V_{tz}}{\partial X} = 0$ because $\vec{V_a}$ is only dependent on $t$.

$$[\vec{X} \quad \vec{Y} \quad \vec{Z}]\begin{bmatrix}\frac{\partial U}{\partial X}\\\frac{\partial Y}{\partial X}\\\frac{\partial W}{\partial X}\end{bmatrix} = [\vec{X} \quad \vec{Y} \quad \vec{Z}]T\begin{bmatrix}\frac{\partial u}{\partial x}\frac{\partial x}{\partial X}+\frac{\partial u}{\partial y}\frac{\partial y}{\partial X}+\frac{\partial u}{\partial z}\frac{\partial z}{\partial X}\\\frac{\partial v}{\partial x}\frac{\partial x}{\partial X}+\frac{\partial v}{\partial y}\frac{\partial y}{\partial X}+\frac{\partial v}{\partial z}\frac{\partial z}{\partial X}\\\frac{\partial w}{\partial x}\frac{\partial x}{\partial X}+\frac{\partial w}{\partial y}\frac{\partial y}{\partial X}+\frac{\partial w}{\partial z}\frac{\partial z}{\partial X}\end{bmatrix} + [\vec{X} \quad \vec{Y} \quad \vec{Z}]\begin{bmatrix}0\\0\\0\end{bmatrix} + [\vec{X} \quad \vec{Y} \quad \vec{Z}]\begin{bmatrix}0\\V_{az}\\-V_{ay}\end{bmatrix} \quad (4.14)$$

Substitute $\frac{\partial x}{\partial X}$, $\frac{\partial y}{\partial X}$ and $\frac{\partial z}{\partial X}$ using Eq.(4.10)

$$[\vec{X} \quad \vec{Y} \quad \vec{Z}]\begin{bmatrix}\frac{\partial U}{\partial X}\\\frac{\partial Y}{\partial X}\\\frac{\partial W}{\partial X}\end{bmatrix} = [\vec{X} \quad \vec{Y} \quad \vec{Z}]T\begin{bmatrix}\frac{\partial u}{\partial x}\cos\alpha_1+\frac{\partial u}{\partial y}\cos\alpha_2+\frac{\partial u}{\partial z}\cos\alpha_3\\\frac{\partial v}{\partial x}\cos\alpha_1+\frac{\partial v}{\partial y}\cos\alpha_2+\frac{\partial v}{\partial z}\cos\alpha_3\\\frac{\partial w}{\partial x}\cos\alpha_1+\frac{\partial w}{\partial y}\cos\alpha_2+\frac{\partial w}{\partial z}\cos\alpha_3\end{bmatrix} + [\vec{X} \quad \vec{Y} \quad \vec{Z}]\begin{bmatrix}0\\0\\0\end{bmatrix} + [\vec{X} \quad \vec{Y} \quad \vec{Z}]\begin{bmatrix}0\\V_{az}\\-V_{ay}\end{bmatrix} \quad (4.15)$$

$$[\vec{X} \quad \vec{Y} \quad \vec{Z}]\begin{bmatrix}\frac{\partial U}{\partial X}\\\frac{\partial Y}{\partial X}\\\frac{\partial W}{\partial X}\end{bmatrix} = [\vec{X} \quad \vec{Y} \quad \vec{Z}]T\begin{bmatrix}\frac{\partial u}{\partial x}&\frac{\partial u}{\partial y}&\frac{\partial u}{\partial z}\\\frac{\partial v}{\partial x}&\frac{\partial v}{\partial y}&\frac{\partial v}{\partial z}\\\frac{\partial w}{\partial x}&\frac{\partial w}{\partial y}&\frac{\partial w}{\partial z}\end{bmatrix}\begin{bmatrix}\cos\alpha_1\\\cos\alpha_2\\\cos\alpha_3\end{bmatrix} + [\vec{X} \quad \vec{Y} \quad \vec{Z}]\begin{bmatrix}0\\V_{az}\\-V_{ay}\end{bmatrix} \quad (4.16)$$

$$[\vec{X} \quad \vec{Y} \quad \vec{Z}]\begin{bmatrix}\frac{\partial U}{\partial X}\\\frac{\partial Y}{\partial X}\\\frac{\partial W}{\partial X}\end{bmatrix} = [\vec{X} \quad \vec{Y} \quad \vec{Z}]\left(T\begin{bmatrix}\frac{\partial u}{\partial x}&\frac{\partial u}{\partial y}&\frac{\partial u}{\partial z}\\\frac{\partial v}{\partial x}&\frac{\partial v}{\partial y}&\frac{\partial v}{\partial z}\\\frac{\partial w}{\partial x}&\frac{\partial w}{\partial y}&\frac{\partial w}{\partial z}\end{bmatrix}\begin{bmatrix}\cos\alpha_1\\\cos\alpha_2\\\cos\alpha_3\end{bmatrix} + \begin{bmatrix}0\\V_{az}\\-V_{ay}\end{bmatrix}\right) \quad (4.17)$$

With the same basis $[\vec{X} \quad \vec{Y} \quad \vec{Z}]$, it can be concluded that the coordinates on the two sides are equal,

$$\begin{bmatrix}\frac{\partial U}{\partial X}\\\frac{\partial V}{\partial X}\\\frac{\partial W}{\partial X}\end{bmatrix} = T\begin{bmatrix}\frac{\partial u}{\partial x}&\frac{\partial u}{\partial y}&\frac{\partial u}{\partial z}\\\frac{\partial v}{\partial x}&\frac{\partial v}{\partial y}&\frac{\partial v}{\partial z}\\\frac{\partial w}{\partial x}&\frac{\partial w}{\partial y}&\frac{\partial w}{\partial z}\end{bmatrix}\begin{bmatrix}\cos\alpha_1\\\cos\alpha_2\\\cos\alpha_3\end{bmatrix} + \begin{bmatrix}0\\V_{az}\\-V_{ay}\end{bmatrix} \quad (4.18)$$

Similarly, we can get

$$\begin{bmatrix}\frac{\partial U}{\partial Y}\\\frac{\partial V}{\partial Y}\\\frac{\partial W}{\partial Y}\end{bmatrix} = T\begin{bmatrix}\frac{\partial u}{\partial x}&\frac{\partial u}{\partial y}&\frac{\partial u}{\partial z}\\\frac{\partial v}{\partial x}&\frac{\partial v}{\partial y}&\frac{\partial v}{\partial z}\\\frac{\partial w}{\partial x}&\frac{\partial w}{\partial y}&\frac{\partial w}{\partial z}\end{bmatrix}\begin{bmatrix}\cos\beta_1\\\cos\beta_2\\\cos\beta_3\end{bmatrix} + \begin{bmatrix}-V_{az}\\0\\V_{ax}\end{bmatrix} \quad (4.19)$$

$$\begin{bmatrix}\frac{\partial U}{\partial Z}\\\frac{\partial V}{\partial Z}\\\frac{\partial W}{\partial Z}\end{bmatrix} = T\begin{bmatrix}\frac{\partial u}{\partial x}&\frac{\partial u}{\partial y}&\frac{\partial u}{\partial z}\\\frac{\partial v}{\partial x}&\frac{\partial v}{\partial y}&\frac{\partial v}{\partial z}\\\frac{\partial w}{\partial x}&\frac{\partial w}{\partial y}&\frac{\partial w}{\partial z}\end{bmatrix}\begin{bmatrix}\cos\gamma_1\\\cos\gamma_2\\\cos\gamma_3\end{bmatrix} + \begin{bmatrix}V_{ay}\\-V_{ax}\\0\end{bmatrix} \quad (4.20)$$

Therefore, the relationship between velocity gradient tensor in the inertial coordinate and the observer's coordinate is

$$\begin{bmatrix} \frac{\partial U}{\partial X} & \frac{\partial U}{\partial Y} & \frac{\partial U}{\partial Z} \\ \frac{\partial V}{\partial X} & \frac{\partial V}{\partial Y} & \frac{\partial V}{\partial Z} \\ \frac{\partial W}{\partial X} & \frac{\partial W}{\partial Y} & \frac{\partial W}{\partial Z} \end{bmatrix} = T \begin{bmatrix} \frac{\partial u}{\partial x} & \frac{\partial u}{\partial y} & \frac{\partial u}{\partial z} \\ \frac{\partial v}{\partial x} & \frac{\partial v}{\partial y} & \frac{\partial v}{\partial z} \\ \frac{\partial w}{\partial x} & \frac{\partial w}{\partial y} & \frac{\partial w}{\partial z} \end{bmatrix} T^T + \begin{bmatrix} 0 & -V_{az} & V_{ay} \\ V_{az} & 0 & -V_{ax} \\ -V_{ay} & V_{ax} & 0 \end{bmatrix} \quad (4.21)$$

where $\vec{V}_a(t) = V_{ax}\vec{X} + V_{ay}\vec{Y} + V_{az}\vec{Z}$ is the angular velocity of the observer measured in the inertial coordinate and T is the transformation matrix from $\vec{x}, \vec{y}, \vec{z}$ to $\vec{X}, \vec{Y}, \vec{Z}$.

It is noted that $V_{ax}, V_{ay}$ and $V_{az}$ are only dependent on $t$, which means they are the same throughout the whole domain. In another word, if we can find out $V_{ax}, V_{ay}$ and $V_{az}$ at one point, these values can be used for the whole domain. $\nabla \vec{v}(\vec{p})$ is the observed result from the observer (measuring instruments), so it is known by the observer. Therefore, the velocity gradient tensor in the inertial coordinate can be found out as long as $V_{ax}, V_{ay}$ and $V_{az}$ at one point is specified.

The method to find $V_{ax}, V_{ay}$ and $V_{az}$ is provided in the following. In the first paragraph of section 3, it is pointed out that people only need to know the result in any one inertial coordinate. So, a special inertial coordinate with the same origin and basis of the observer's coordinate can be selected to make the analysis simpler.

***Theorem 1***: If $\vec{P}$ is a point with zero vorticity in the inertial XYZ coordinate with the same origin and basis of the observer's coordinate, then $V_{ax}\vec{x} + V_{ay}\vec{y} + V_{az}\vec{z} = -\frac{1}{2}(\omega_x\vec{x} + \omega_y\vec{y} + \omega_z\vec{z})$ where $\omega_x\vec{x} + \omega_y\vec{y} + \omega_z\vec{z}$ is the vorticity of the corresponding $\vec{p}$ under the observer's coordinate.

***Proof***: According to the Cauchy-Stokes decomposition, the $\vec{V}(\vec{P})$ (measured in the inertial coordinate) can be decomposed to symmetric $A$ and antisymmetric part $B$,

$$\nabla \vec{V} = A + B \quad (4.22)$$

where

$$A = \frac{1}{2}\left[\nabla \vec{V} + (\nabla \vec{V})^T\right] \qquad B = \frac{1}{2}\left[\nabla \vec{V} - (\nabla \vec{V})^T\right] \quad (4.23)$$

B is a zero matrix because the vorticity is zero at the point.
Substitute $\nabla \vec{V}$ with Eq.(4.23) in Eq.(4.22)

$$A + B = T\nabla \vec{v} T^T + \begin{bmatrix} 0 & -V_{az} & V_{ay} \\ V_{az} & 0 & -V_{ax} \\ -V_{ay} & V_{ax} & 0 \end{bmatrix} \quad (4.24)$$

$\nabla \vec{v}$ can be solved from Eq. (4.24)

$$\nabla \vec{v} = T^T A T - T^T \begin{bmatrix} 0 & -V_{az} & V_{ay} \\ V_{az} & 0 & -V_{ax} \\ -V_{ay} & V_{ax} & 0 \end{bmatrix} T \quad (4.25)$$

Here, T is an identity matrix because the inertial coordinate selected is with the same origin and basis of the observer. Thus,

$$T = \begin{bmatrix} 1 & 0 & 0 \\ 0 & 1 & 0 \\ 0 & 0 & 1 \end{bmatrix} \quad (4.26)$$

Then,

$$\nabla \vec{v} = A - \begin{bmatrix} 0 & -V_{az} & V_{ay} \\ V_{az} & 0 & -V_{ax} \\ -V_{ay} & V_{ax} & 0 \end{bmatrix} = A + \begin{bmatrix} 0 & V_{az} & -V_{ay} \\ -V_{az} & 0 & V_{ax} \\ V_{ay} & -V_{ax} & 0 \end{bmatrix} \qquad (4.27)$$

Obviously, $\begin{bmatrix} 0 & V_{az} & -V_{ay} \\ -V_{az} & 0 & V_{ax} \\ V_{ay} & -V_{ax} & 0 \end{bmatrix}$ is the vorticity matrix measured in the observer's coordinate since $A$ is a symmetric matrix. Therefore,

$$V_{ax}\vec{x} + V_{ay}\vec{y} + V_{az}\vec{z} = -\frac{1}{2}(\omega_x\vec{x} + \omega_y\vec{y} + \omega_z\vec{z}) \qquad (4.28)$$

Since the basis on the two sides are the same, we have

$$(V_{ax}, V_{ay}, V_{az}) = -\frac{1}{2}(\omega_x, \omega_y, \omega_z) \qquad (4.29)$$

where $\omega_x, \omega_y, \omega_z$ are the vorticity components of $\nabla \vec{v}$. The theorem is proved.

The velocity gradient tensors of all points in the inertial coordinate (objective velocity gradient tensor) can be obtained by

$$\nabla \vec{V} = \nabla \vec{v} + \begin{bmatrix} 0 & -V_{az} & V_{ay} \\ V_{az} & 0 & -V_{ax} \\ -V_{ay} & V_{ax} & 0 \end{bmatrix} \qquad (4.30)$$

where $\nabla \vec{v}$ comes from the observer's measurements and $(V_{ax}, V_{ay}, V_{az})$ calculated from Eq. (4.29). Eq. (4.30) gives the objective velocity gradient tensor result since velocity gradient tensor is a Galilean invariant concept.

The velocity in an inertial coordinate can be obtained based on the Eq. (4.1). A special inertial coordinate is selected such that it is with the same origin and same basis of the observer, and it is doing uniform rectilinear motion so that $\vec{V}_t(t) = 0$. (The observer's translation velocity is 0 measured in the inertial coordinate). For such a special inertial coordinate, $\vec{V}_t(t) = 0$; $\vec{P} = \vec{p}$ since the two coordinates share the same origin and basis; $\vec{V}_a(t)$ can be figured out from Eq. (4.29); $\vec{v}(\vec{p})$ comes from the measurement of the observer. Thus, all the terms on the right-hand side of Eq. (4.1) are known. And the simplified formula becomes

$$\vec{V}(\vec{P}) = \vec{v}(\vec{p}) + \vec{V}_a(t) \times \vec{p} \qquad (4.31)$$

It is emphasized that Eq. (4.31) does not give objective velocity results since velocity is not Galilean invariant but the velocity distribution in a special inertial coordinate can be known.

Two strategies can be derived based on the velocity or velocity gradient tensor in a special inertial coordinate.

**Strategy 1**(based on velocity gradient tensor):
1. Pick a point with zero vorticity. The point can be selected based on physical properties of the flow e.g., points in the inviscid region.
2. Calculate vorticity at the selected point. Then $(V_{ax}, V_{ay}, V_{az}) = -\frac{1}{2}(\omega_x, \omega_y, \omega_z)$.
3. The velocity gradient tensor of all points in an inertial coordinate can be obtained from

$$\nabla \vec{V} = \nabla \vec{v} + \begin{bmatrix} 0 & -V_{az} & V_{ay} \\ V_{az} & 0 & -V_{ax} \\ -V_{ay} & V_{ax} & 0 \end{bmatrix} \quad (4.32)$$

4. Calculate Liutex from $\nabla \vec{V}$.

Any other Galilean invariant vortex identification methods based on the velocity gradient tensor can also follow the Strategy 1 to get objective results. The new Liutex calculated from $\nabla \vec{V}$ is objective or defined as $\vec{L}_{ob}$

**Strategy 2** (based on velocity):
1. Pick a point with zero vorticity. The point can be selected based on physical property of the flow e.g., points in the inviscid region.
2. Calculate vorticity at the selected point. Then $(V_{ax}, V_{ay}, V_{az}) = -\frac{1}{2}(\omega_x, \omega_y, \omega_z)$.
3. The velocity of all points in an inertial coordinate can be obtained from
$$\vec{V}(\vec{P}) = \vec{v}(\vec{p}) + \vec{V}_a(t) \times \vec{p} \quad (4.33)$$
4. Calculate objective velocity gradient tensor.
5. Calculate new Liutex or $\vec{L}_{ob}$.

Any other Galilean invariant vortex identification methods based on the velocity can also follow the above Strategy 2 to get objective Liutex or $\vec{L}_{ob}$.

## 5. Numerical Tests

Two numerical simulation results are used to test the provided strategies. The first case is the direct numerical simulation (DNS) of the flat plat boundary layer transition (Wang et al. 2017). Its mesh size is $1920 \times 128 \times 241$ in streamwise(x), spanwise(y) and normal(z) directions respectively. The properties of the flow are shown in Table 1. $M_\infty$ is the Mach number; Re represents Reynolds number; $x_{in}$ is the distance between leading edge of flat plate and upstream boundary of computational domain; $Lx$ and $Ly$ are the lengths of computational domain along $x$ and $y$ directions respectively; $Lz_{in}$ means the height at inflow boundary; $T_w$ and $T_\infty$ represent the temperature of wall and free stream respectively.

| $M_\infty$ | Re | $x_{in}$ | $Lx$ | $Ly$ | $Lz_{in}$ | $T_w$ | $T_\infty$ |
|---|---|---|---|---|---|---|---|
| 0.5 | 1000 | $300.79\delta_{in}$ | $798.03\delta_{in}$ | $22\delta_{in}$ | $40\delta_{in}$ | 273.15K | 273.15K |

Table 1 Flow Properties

The graph of the original DNS result with Liutex=0.07 iso-surface is shown in Fig. 2.

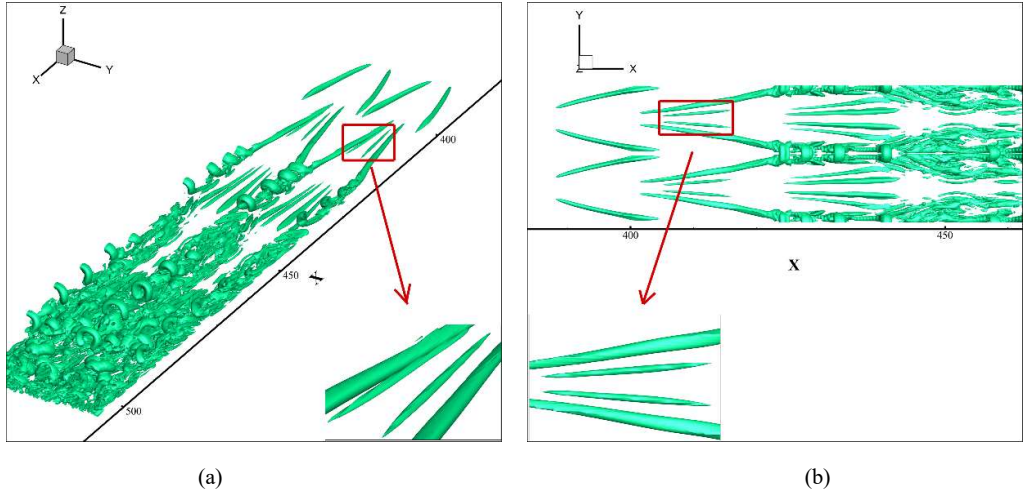

(a) (b)

Fig.2 Vortex structure in the inertial original coordinate with Liutex=0.07 iso-surface (a) overall (b) from the top

An artificial angular velocity (observer's angular velocity) $\vec{V_a} = (0.009, 0.008, 0.007)$ is applied. The vortex structure in the observer's view is shown in Fig.3. It can be seen obviously that the vortex structures in these two different coordinates are different. In the original inertial coordinate, the legs look almost symmetric while in the observer's(non-inertial) coordinate, the legs lose the symmetry.

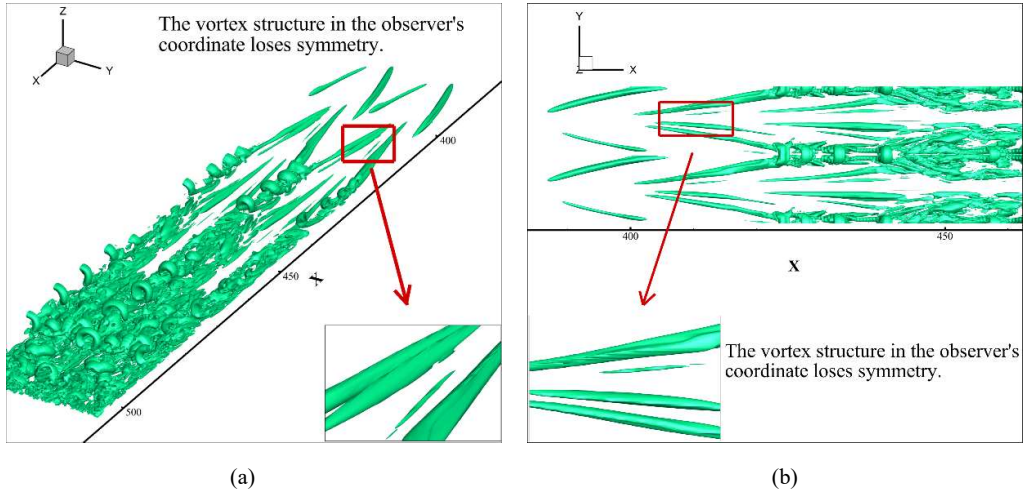

(a) (b)

Fig.3 Vortex structure in the observer's coordinate with Liutex=0.07 iso-surface (a) overall (b) from the top

The vorticity at the point (in the inviscid region) x=456.38, y=20.62, z=13.36 measured in the observer's coordinate is (-0.0179999905974, -0.0159999989568, -0.0140000715749). From Eq. (4.29), the estimated angular velocity of the observer is

$$\begin{bmatrix} V_{ax} \\ V_{ay} \\ V_{az} \end{bmatrix} = -\frac{1}{2} \begin{bmatrix} -0.0179999905974 \\ -0.0159999989568 \\ -0.0140000715749 \end{bmatrix} = \begin{bmatrix} 0.0089999952986874916 \\ 0.0079999994783875872 \\ 0.0070000357874608187 \end{bmatrix} \quad (5.1)$$

The estimated angular velocity is very close to real angular velocity.

Use Eq. (4.30) to calculate the objective velocity gradient tensor result,

$$\nabla \vec{V} = \nabla \vec{v} + \begin{bmatrix} 0 & -0.0070000357874608187 & 0.0079999994783875872 \\ 0.0070000357874608187 & 0 & -0.0089999952986874916 \\ -0.0079999994783875872 & 0.0089999952986874916 & 0 \end{bmatrix} \quad (5.2)$$

Calculate Liutex based on the objective velocity gradient tensor result and the graph of the vortex structure is shown in Fig.4.

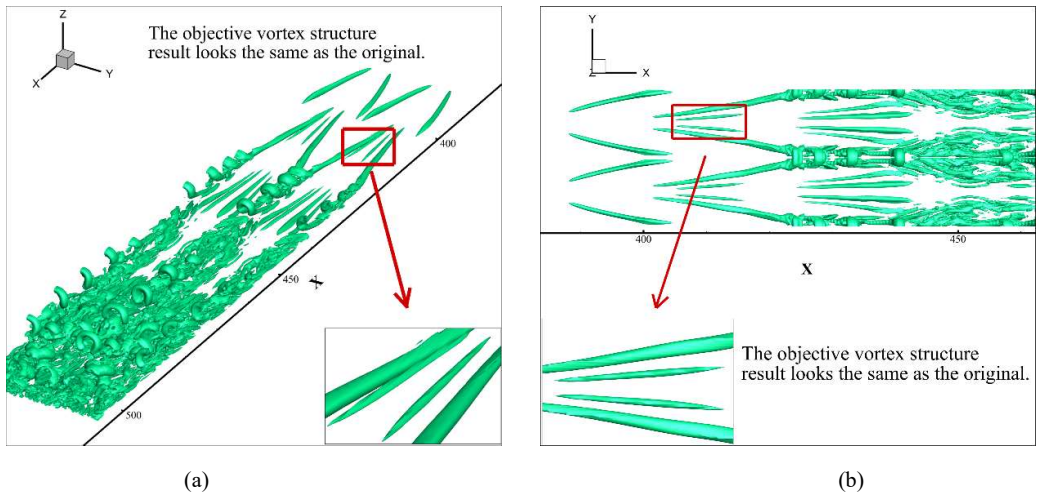

(a)　　　　　　　　　　　　　　　　　　　(b)

Fig.4 Objective Vortex with Liutex=0.07 iso-surface (a) overall (b) from the top

In the original coordinate, the vortex structure has symmetry in the early stage and the lengths of the two legs look the same in the locally enlarged figures. While the observation by the non-inertial observer shows that the symmetry loses in the early stage and the lengths of the two legs have obvious difference. After getting the objective Liutex result using the strategy proposed in this paper, there is no difference can be seen by eyes between the original vortex structure and the objective Liutex result which shows the effectiveness of the strategy. The new objective Liutex or $\vec{L}_{ob}$ is objective to any moving and rotating frames or observers.

To test the strategy quantitively, the Liutex magnitude in the original coordinate and the objective results are compare at the 12 selected points (see Fig.5). The result is shown in Table 2.

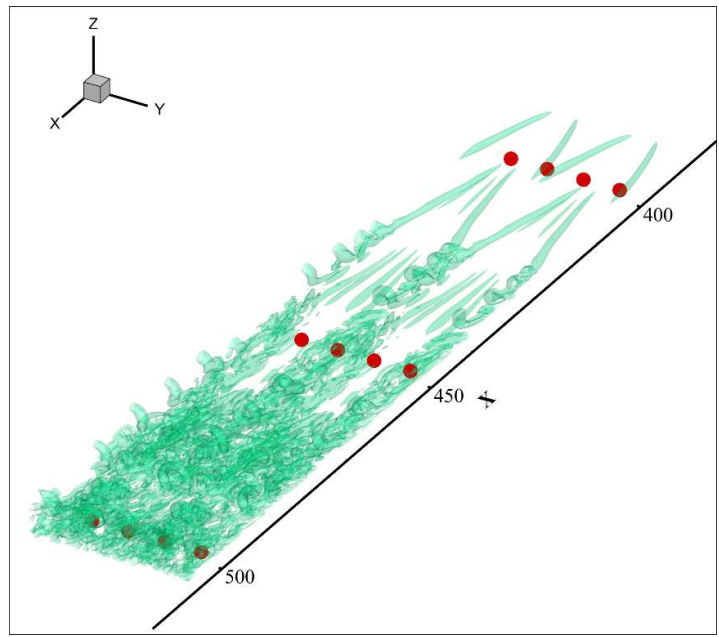

Fig 5 Locations of the selected points.

It is easy to find that Liutex magnitudes in the original coordinate are the same with the objective Liutex results calculated from the observation of a non-inertial observer. This means the magnitude calculated in a non-inertial coordinate is almost exactly same as one in the original inertial frame. The new Liutex or $\vec{L}_{ob}$ is objective, which is independent of observers or any non-inertial coordinates.

|  | Original | Objective | Magnitude Error |
|---|---|---|---|
| x=400 y=5 z=0.5 | (-0.00485488778985407061, 0.00360010337248459125, -0.000210512257145231696) | (-0.00485488778985407061, 0.00360010337248459125, -0.000210512257145231696) | 0% |
| x=400 y=10 z=0.5 | (0.0136990846926978941, 0.00868670662529319113, -0.000319871832422033214) | (0.0136990846926978941, 0.00868670662529319113, -0.000319871832422033214) | 0% |
| x=400 y=15 z=0.5 | (0.0049990121508986839, 0.00102997258119983661, 0.000445101127617844291) | (0.0049990121508986839, 0.00102997258119983661, 0.000445101127617844291) | 0% |
| x=400 y=20 z=0.5 | (0.0106496344770713146, 0.00339052410928388165, -0.000728406832035471247) | (0.0106496344770713146, 0.00339052410928388165, -0.000728406832035471247) | 0% |
| x=450 y=5 z=0.5 | (0.0331574530100377032, 0.00175911459627213105, 0.00181067498447680603) | (0.0331574530100377032, 0.00175911459627213105, 0.00181067498447680603) | 0% |
| x=450 y=10 z=0.5 | (0.00770432455445506457, 0.00344814790921462149, 0.000234320590533778771) | (0.00770432455445506457, 0.00344814790921462149, 0.000234320590533778771) | 0% |
| x=450 y=15 z=0.5 | (-0.0691773294397066418, -0.00654341117059082083, -0.00584994836046827654) | (-0.0691773294397066418, -0.00654341117059082083, -0.00584994836046827654) | 0% |
| x=450 y=20 z=0.5 | (-0.0170553536109858614, -0.00356296277622985278, 0.00124746325046161443) | (-0.0170553536109858614, -0.00356296277622985278, 0.00124746325046161443) | 0% |
| x=500 y=5 z=0.5 | (0.196342488744981392, -0.00449951934699422719, 0.0615913155993694852) | (0.196342488744981392, -0.00449951934699422719, 0.0615913155993694852) | 0% |
| x=500 y=10 z=0.5 | (0.00671670231952755693, 0.00400440747481066624, 0.00102951830811361502) | (0.00671670231952755693, 0.00400440747481066624, 0.00102951830811361502) | 0% |
| x=500 y=15 z=0.5 | (-0.0301061021760056988, 0.0298032729923758608, -0.0194201813823916963) | (-0.0301061021760056988, 0.0298032729923758608, -0.0194201813823916963) | 0% |
| x=500 y=20 z=0.5 | (0.309306801381793506, 0.013841870044603008, 0.0802540370146049425) | (0.309306801381793506, 0.013841870044603008, 0.0802540370146049425) | 0% |

Table 2 Original and Objective Liutex magnitude at the selected points with their relative errors

Another simulation of Shock wave boundary-layer interaction (SBLI) is used to test the strategy as well. The grid size is $137 \times 192 \times 1600$ (spanwise×normal×streamwise). The dimension of the ramp is shown in Fig.6. Fully developed turbulence is generated in the front of the ramp. The oblique shock wave causes a larger area which has supersonic flow. More details can be found in Yan (Yan et al. 2017)

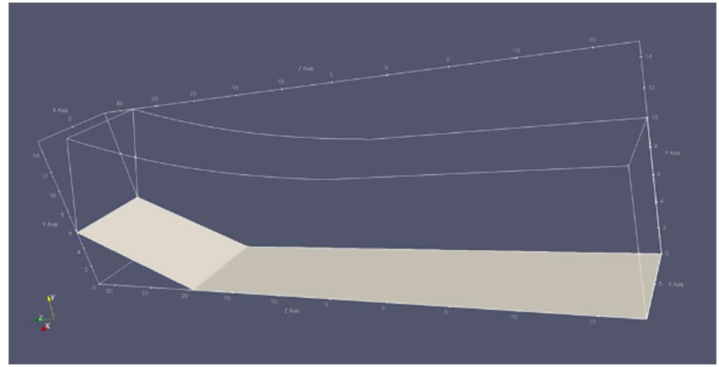

Fig.6 Dimension of the ramp

The original graph with iso-surface Liutex=0.07 is shown in Fig. 7.

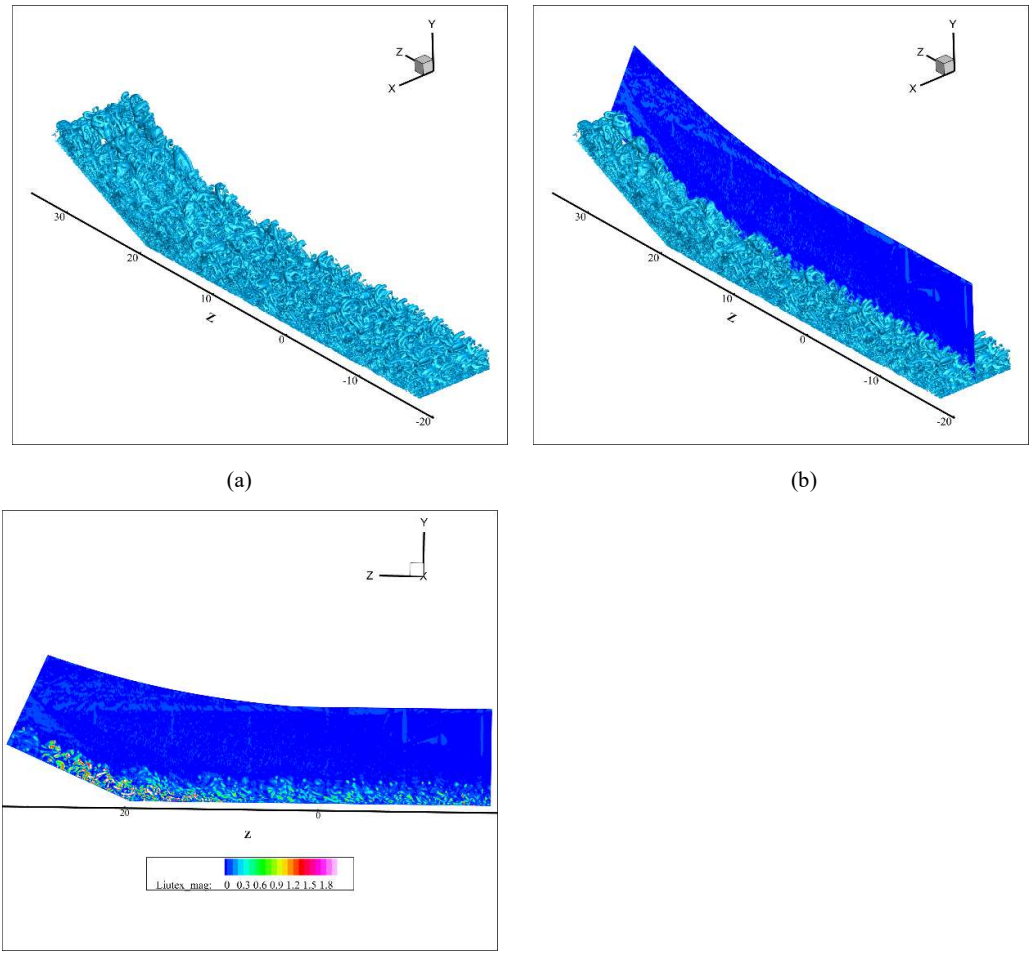

(a)　　　　　　　　　　　　　　　　(b)

(c)

Fig. 7 The vortex structure in the original coordinate (a) with iso-surface Liutex=0.1 (b) with iso-surface Liutex=0.1 and the slice (c) central section only.

Then an artificial angular velocity $\vec{V}_a = (0.03, 0.04, 0.05)$ of the observer is applied. The graph of the observer's view can be seen in Fig.8. In the observer's (non-inertial) coordinate, an oblique laminar Liutex iso-surface appears at the position A and a horizontal laminar Liutex iso-surface appears at the position B. These differences compared with Fig. 7 are caused by the angular velocity of the observer.

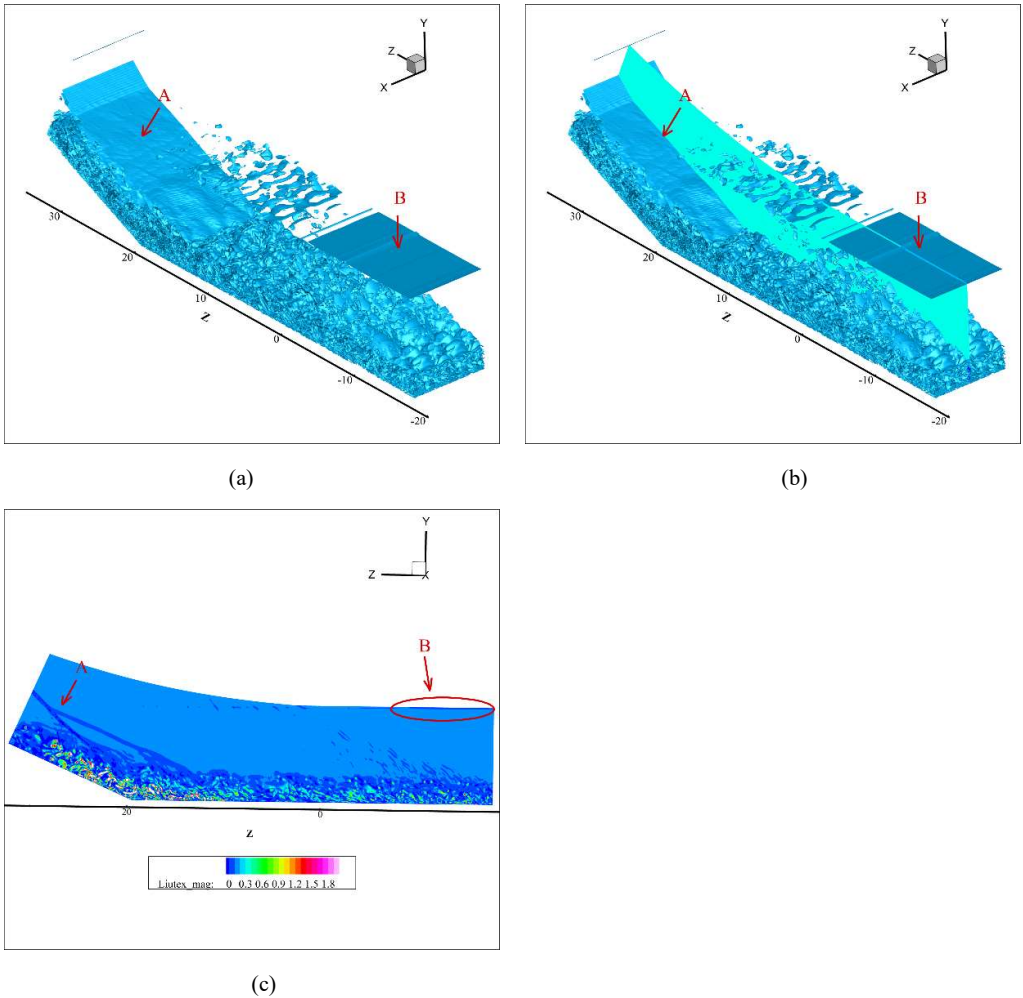

(a)          (b)

(c)

Fig. 8 The vortex structure in the observer's coordinate (a) with iso-surface Liutex=0.1, (b) with iso-surface Liutex=0.1 and the slice, (c) central section only.

Use the point x=3.75, y=4.45, z=-14.20 which is located in the inviscid region to calculate the estimated observer's angular velocity. Its vorticity is (-0.060000000596234448, -0.079999998978732584, -0.099999999320150551) which implies estimated observer's angular velocity is

$$\begin{bmatrix} V_{ax} \\ V_{ay} \\ V_{az} \end{bmatrix} = -\frac{1}{2}\begin{bmatrix} -0.060000000596234448 \\ -0.079999998978732584 \\ -0.099999999320150551 \end{bmatrix} = \begin{bmatrix} 0.030000000298117224 \\ 0.039999999489366292 \\ 0.049999999660075276 \end{bmatrix} \quad (5.3)$$

The estimated angular velocity is very close to the real angular velocity of the observer. Apply Eq. (4.30) to calculate the velocity gradient tensor field in an inertial coordinate.

$$\nabla \vec{V} = \nabla \vec{v} + \begin{bmatrix} 0 & -0.049999999660075276 & 0.039999999489366292 \\ 0.049999999660075276 & 0 & -0.030000000298117224 \\ -0.039999999489366292 & 0.030000000298117224 & 0 \end{bmatrix} \quad (5.4)$$

Objective Liutex results can be calculated based on the velocity gradient tensor field. The graph of objective Liutex is shown in Fig. 9.

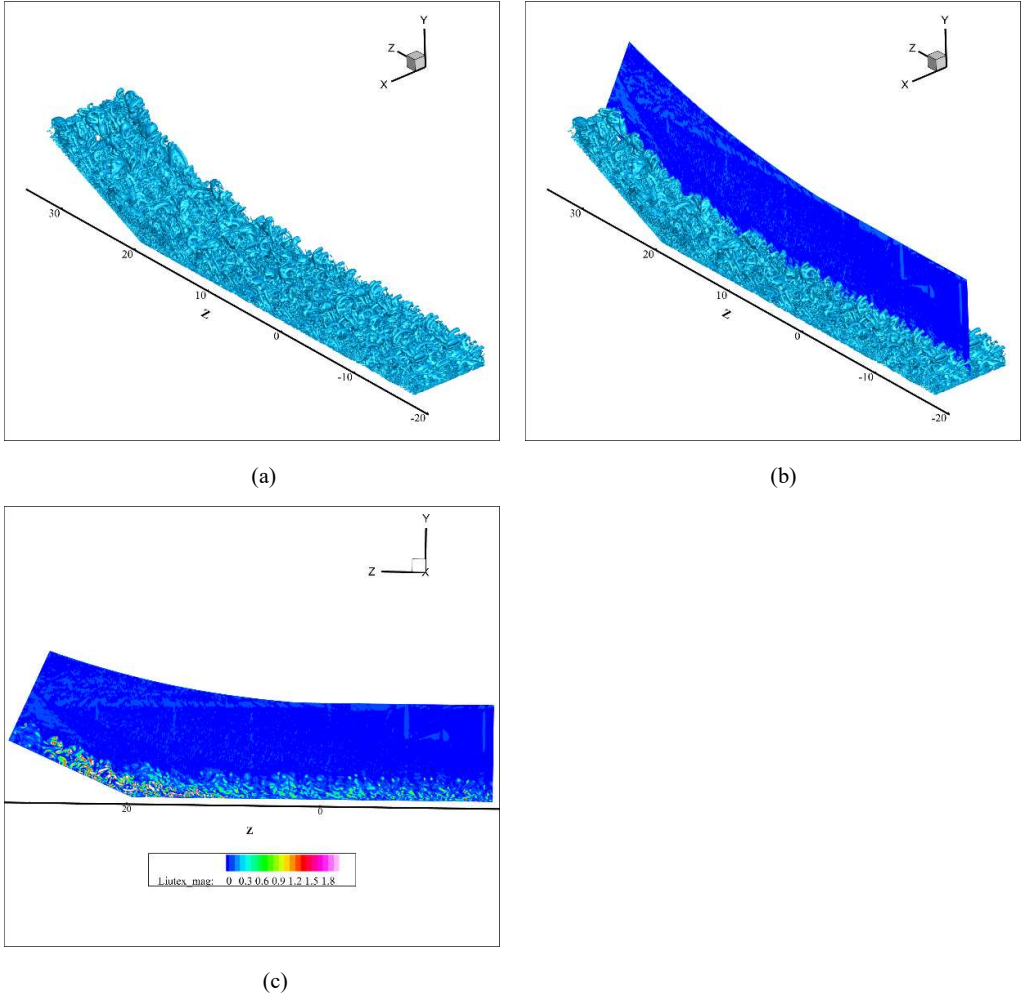

(a)  (b)

(c)

Fig. 9 The objective vortex structure (a) with iso-surface Liutex=0.1 (b) with iso-surface Liutex=0.1 and the slice (c) central section only.

20 points are selected to compare the original Liutex and objective Liutex quantitively. The distribution of the points can be found in Fig.10. The data is shown in table 3.

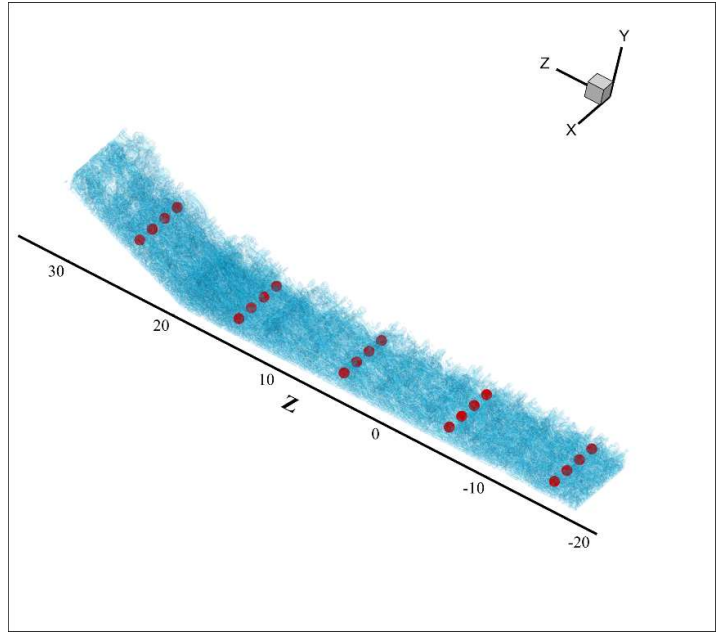

Fig. 10 The locations of the selected points.

|  | Original | Objective | Magnitude Error |
|---|---|---|---|
| x=-3  y=1  z=-15 | (0.000395086478584978943, 0.00006939812165074128, -0.000201683061626090057) | (0.000395086506897557462, 0.00006.93981269330667071, -0.000201683077093950988) | $7.278 \times 10^{-8}$ |
| x=-1  y=1  z=-15 | (0.0534089728215503615, -0.0415490918899853595, -0.0544567766272447892) | (0.0534089736161507192, -0.0415490925670093936, -0.0544567775230459447) | $1.550 \times 10^{-8}$ |
| x=1  y=1  z=-15 | (0.0462985682629876716, 0.0132302896528897918, 0.00555756114120387144) | (0.0462985676765630544, 0.0132302895094048821, 0.00555756104455938475) | $1.261 \times 10^{-8}$ |
| x=3  y=1  z=-15 | (0.0303965244418970097, 0.0318027599980555201, 0.0890728180132685649) | (0.0303965242711929581, 0.0318027599692987799, 0.0890728174070746365) | $6.959 \times 10^{-9}$ |
| x=-3  y=1  z=-5 | (0.00110155401644547717, 0.000420114198309581291, 0.00102410845209206011) | (0.00110155402917517728, 0.000420114199750452476, 0.00102410846763573865) | $1.253 \times 10^{-8}$ |
| x=-1  y=1  z=-5 | (0,0,0) | (0,0,0) | 0 |
| x=1  y=1  z=-5 | (0.000607582613439754788, 0.00019885919938753574, -0.000141279185053922016) | (0.000607582669150267062, 0.00019885921998579522, -0.000141279199846237175) | $1.251 \times 10^{-7}$ |
| x=3  y=1  z=-5 | (0.00321473487113252019, -0.000942848196755211227, -0.00104502321659029340) | (0.00321473502862204118, -0.000942848245299535712, -0.00104502325974545731) | $4.850 \times 10^{-8}$ |
| x=-3 y=1 z=5 | (0.0572182385697277071, -0.0192497849334081969, 0.04493151476987714) | (0.0572182387498169701, -0.0192497851028640456, 0.0449315149463732078) | $3.805 \times 10^{-9}$ |
| x=-1 y=1 z=5 | (-0.125442285461300623, -0.0633555689287845503, -0.0678009895950665853) | (-0.125442286004231102, -0.06335557001931795, -0.0678009902089121846) | $7.311 \times 10^{-9}$ |
| x=1 y=1 z=5 | (0.0108345272073865143, 0.00380005166892837063, -0.00364433390509177023) | (0.0108345272034985809, 0.0038000516673340045, -0.00364433396990331525) | $6.027 \times 10^{-10}$ |
| x=3 y=1 z=5 | (0.0302051846692774975, 0.0285076696411926955, 0.0860330069183567331) | (0.0302051844159486396, 0.0285076692881624238, 0.0860330064586181942) | $6.125 \times 10^{-9}$ |
| x=-3 y=1 z=15 | (0.354237139784671029, -0.139701571405804359, -0.176865016115715951) | (0.354237140598927691, -0.139701572006825009, -0.176865016974736361) | $2.877 \times 10^{-9}$ |
| x=-1 y=1 z=15 | (0.275868341904399239, -0.478522662560607992, -0.396642384808442272) | (0.275868342540467937, -0.478522663541999183, -0.396642385523431173) | $1.996 \times 10^{-9}$ |
| x=1 y=1 z=15 | (0,0,0) | (0,0,0) | 0 |
| x=3 y=1 z=15 | (-0.00102306813140086526, 0.003708975488590378, 0.00417541074550020503) | (-0.00102306810563998507, 0.00370897538931861835, 0.00417541066434626382) | $1.921 \times 10^{-8}$ |
| x=-3 y=3.5 z=25 | (0.0113134027097314127, 0.0236499022863370968, 0.0469846361114432468) | (0.0113134026672318914, 0.0236499021847234314, 0.0469846359465392466) | $3.673 \times 10^{-9}$ |
| x=-1 y=3.5 z=25 | (0,0,0) | (0,0,0) | 0 |
| x=1 y=3.5 z=25 | (0,0,0) | (0,0,0) | 0 |
| x=3 y=3.5 z=25 | (0,0,0) | (0,0,0) | 0 |

Table 3 Original and Objective Liutex magnitude at the selected points with their relative errors

The error between the objective Liutex in a rotating frame and the original Liutex in an inertial frame is very small, at most at the order of $10^{-7}$.

In summary, the proposed strategies in this paper exhibit very good ability to detect the angular velocity of the observer (non-inertial), and to transfer the non-inertial observation to the results under an inertial coordinate from the above two numerical tests.

6. Conclusion
(1) Liutex given by Eq. (3.1) is Galilean invariant, but not objective

(2) We must first find an inviscid point where vorticity=0 as a reference point or $\vec{P}$, and then find its velocity or velocity gradient.

(3) Applying Eq. (4.23) and (4.33) which involve the vorticity of the reference point $\vec{P}$, we can get new velocity $\vec{V}$ and new velocity gradient $\nabla\vec{V}$

(4) Use new $\vec{V}$ or $\nabla\vec{V}$ and Eq. (3.1), we can get $\vec{L}_{ob}$ which is the objective Liutex

(5) Objective Liutex $\vec{L}_{ob}$ can well describe vortex structure and keeps invariant in any non-inertial coordinate

(6) Two DNS and LES examples show $\vec{L}_{ob}$ is objective and the error generated by the proposed method are ignorable.

(7) The method proposed in this paper is simple and straight forward to get objective vortex structure by new objective Liutex.

7. Data availability

The data that supports the findings of this study are available from the corresponding author upon reasonable request.

Acknowledgement:

The authors acknowledge the financial support from UTA Mathematics Department and are grateful to Texas Advanced Computation Center (TACC) for providing computing resources.

Reference:

Chong, M. S., A. E. Perry, and B. J. Cantwell. 1990. 'A general classification of three-dimensional flow fields', *Physics of Fluids A: Fluid Dynamics*, 2: 765-77.

Dong, Xiangrui, and Chaoqun Liu. 2021. 'Micro-Ramp Wake Structures Identified by Liutex.' in, *Liutex and Third Generation of Vortex Definition and Identification* (Springer).

Günther, Tobias, Markus Gross, and Holger Theisel. 2017. 'Generic objective vortices for flow visualization', *ACM Transactions on Graphics*, 36: 1-11.

Guo, Yan-ang, Xiang-rui Dong, Xiao-shu Cai, and Wu Zhou. 2020. 'Experimental study on dynamic mechanism of vortex evolution in a turbulent boundary layer of low Reynolds number', *Journal of Hydrodynamics*, 32: 807-19.

Gurtin, Morton E. 1982. *An introduction to continuum mechanics* (Academic press).

Gurtin, Morton E, Eliot Fried, and Lallit Anand. 2010. *The mechanics and thermodynamics of continua* (Cambridge University Press).

Hadwiger, M., M. Mlejnek, T. Theusl, and P. Rautek. 2018. 'Time-Dependent Flow seen through Approximate Observer Killing Fields', *IEEE Trans Vis Comput Graph*.

Haller, G. 2005. 'An objective definition of a vortex', *Journal of Fluid Mechanics*, 525: 1-26.

Hunt, Julian CR, Alan A Wray, and Parviz Moin. 1988. 'Eddies, streams, and convergence zones in turbulent flows', *Studying Turbulence Using Numerical Simulation Databases, 2. Proceedings of the 1988 Summer Program*.

Jeong, Jinhee, and Fazle Hussain. 1995. 'On the identification of a vortex', *Journal of Fluid Mechanics*, 285.

Kreilos, Tobias, Stefan Zammert, and Bruno Eckhardt. 2014. 'Comoving frames and symmetry-related motions in parallel shear flows', *Journal of Fluid Mechanics*, 751: 685-97.

Landau, Lev Davidovich. 2013. *The classical theory of fields* (Elsevier).

Liang, Yunzhi, Zuti Zhang, Huaiyu Cheng, and Xinping Long. 2022. "Numerical investigation on the interaction between cavitation and vortex in a Venturi tube based on Liutex method." In *Journal of Physics: Conference Series*, 012012. IOP Publishing.

Liu, Chaoqun, Yisheng Gao, Shuling Tian, and Xiangrui Dong. 2018. 'Rortex—A new vortex vector definition and vorticity tensor and vector decompositions', *Physics of Fluids*, 30.

Liu, Jianming, Yisheng Gao, and Chaoqun Liu. 2019. 'An objective version of the Rortex vector for vortex identification', *Physics of Fluids*, 31: 065112.

Liu, Yiqian Wang; Yisheng Gao; Chaoqun. 2018. 'Letter: Galilean invariance of Rortex', *Physics of Fluids*, 30.

Lugt, Hans J. 1979. 'The dilemma of defining a vortex.' in, *Recent developments in theoretical and experimental fluid mechanics* (Springer).

Martins, Ramon S., Anselmo Soeiro Pereira, Gilmar Mompean, Laurent Thais, and Roney Leon Thompson. 2016. 'An objective perspective for classic flow classification criteria', *Comptes Rendus Mécanique*, 344: 52-59.

Mellibovsky, F., and B. Eckhardt. 2012. 'From travelling waves to mild chaos: a supercritical bifurcation cascade in pipe flow', *Journal of Fluid Mechanics*, 709: 149-90.

Robinson, S. K. 1991. 'COHERENT MOTIONS IN THE TURBULENT BOUNDARY-LAYER', *Annual Review of Fluid Mechanics*, 23: 601-39.

Rojo, I. B., and T. Gunther. 2020. 'Vector Field Topology of Time-Dependent Flows in a Steady Reference Frame', *IEEE Trans Vis Comput Graph*, 26: 280-90.

Shrestha, Pushpa, Charles Nottage, Yifei Yu, Oscar Alvarez, and Chaoqun Liu. 2021. 'Stretching and shearing contamination analysis for Liutex and other vortex identification methods', *Advances in Aerodynamics*, 3.

Waleffe, Fabian. 2001. 'Exact coherent structures in channel flow', *Journal of Fluid Mechanics*, 435: 93-102.

Wang, Yiqian, Yong Yang, Guang Yang, and Chaoqun Liu. 2017. 'DNS Study on Vortex and Vorticity in Late Boundary Layer Transition', *Communications in Computational Physics*, 22: 441-59.

Wang, Yu-fan, Wei-hao Zhang, Xia Cao, and Hong-kai Yang. 2019. 'The applicability of vortex identification methods for complex vortex structures in axial turbine rotor passages', *Journal of Hydrodynamics*, 31: 700-07.

Yan, Y., L. Chen, Q. Li, and C. Liu. 2017. 'Numerical study of micro-ramp vortex generator for supersonic ramp flow control at Mach 2.5', *Shock Waves*, 27: 79-96.

Yu, Yifei, Pushpa Shrestha, Oscar Alvarez, Charles Nottage, and Chaoqun Liu. 2021. 'Investigation of correlation between vorticity, Q, λci, λ2, Δ and Liutex', *Computers & Fluids*, 225.

Zhao, Min-sheng, Wei-wen Zhao, De-cheng Wan, and Yi-qian Wang. 2021. 'Applications of Liutex-based force field models for cavitation simulation', *Journal of Hydrodynamics*, 33: 488-93.

Zhao, Wei-wen, Jian-hua Wang, and De-cheng Wan. 2020. 'Vortex identification methods in marine hydrodynamics', *Journal of Hydrodynamics*, 32: 286-95.

Zhou, J., R. J. Adrian, S. Balachandar, and T. M. Kendall. 1999. 'Mechanisms for generating coherent packets of hairpin vortices in channel flow', *Journal of Fluid Mechanics*, 387: 353-96.